\documentclass[conference,english]{IEEEtran}
\IEEEoverridecommandlockouts
\usepackage{cite}
\usepackage{amsmath,amssymb,amsfonts}
\usepackage{algorithmic}
\usepackage{graphicx}
\usepackage{textcomp}
\usepackage{calc,pifont} 
\usepackage{lipsum, babel}
\usepackage{graphics} 
\usepackage{epsfig} 
\usepackage{mathptmx} 
\usepackage{times} 
\usepackage{amsthm}
\usepackage{amssymb, amsmath}
\usepackage{epstopdf} 
\usepackage{cite}
\usepackage{thmtools}
\usepackage{etoolbox}
\usepackage{subeqnarray}
\usepackage{cases}
\usepackage{wrapfig}
\usepackage{graphicx}
\usepackage{accents}

\DeclareMathOperator{\tr}{tr} %
\DeclareMathOperator{\im}{Im} %
\DeclareMathOperator{\re}{Re} %

\newtheorem{rmk}{Remark}
\usepackage{tabularx}
\allowdisplaybreaks
\begin{document}
\title{A Derivation of Moment Evolution Equations for Linear Open Quantum Systems* \thanks{\textsuperscript{*}The material presented here has been partially presented in thesis form \cite{M17:thesis}. This work was supported by the 111 Project (B17048), the Australian Research Council
(ARC) under grant FL110100020, the Air Force Office of Scientific Research
(AFOSR), under agreement number FA2386-16-1-4065, and the Australian
Academy of Science.}}

\author{\IEEEauthorblockN{Shan Ma}
\IEEEauthorblockA{\textit{School of Information Science and Engineering} \\
\textit{Central South University}\\
Changsha 410083, China \\
\textit{School of Engineering and Information Technology} \\
\textit{UNSW Canberra}\\
Canberra ACT 2600, Australia \\
shanma.adfa@gmail.com}
\and
\IEEEauthorblockN{Matthew J. Woolley}
\IEEEauthorblockA{\textit{School of Engineering and IT } \\
\textit{UNSW Canberra}\\
Canberra ACT 2600, Australia \\
m.woolley@adfa.edu.au}
\and 
\IEEEauthorblockN{Ian R. Petersen}
\IEEEauthorblockA{\textit{Research School of Engineering} \\
\textit{The Australian National University}\\
 Canberra ACT 2601， Australia\\
i.r.petersen@gmail.com}
}

\maketitle

\begin{abstract}
Given a linear open quantum system which is described by a Lindblad master equation, we detail the calculation of the moment evolution equations from this master equation. 
We stress that the moment evolution equations are well-known, but their explicit derivation from the master equation cannot be found in the literature to the best of our knowledge, and so we provide this derivation for the interested reader. 
\end{abstract}


\section{Introduction}
Quantum systems unavoidably interact with their surrounding environments, and such quantum systems are referred to as open quantum systems. The dynamics  of open quantum systems can be classified into two categories, Markovian and non-Markovian regimes, depending on whether the system is \emph{weakly} coupled to a \emph{memoryless} environment~\cite{BP02:book,GZ00:book}. For open quantum systems coupled weakly to a memoryless
environment, a well-established treatment known as a \emph{Markovian Lindblad master equation} can be used to approximate the time evolution of such systems~\cite{GKS76:jmp,L76:cmp,GZ00:book,BP02:book}. Typical examples of such systems include 
quantum optical systems~\cite{WM08:book}. 

Let us consider a continuous-variable open quantum system with $ N$
degrees of freedom, the time evolution of which is described by the following Markovian Lindblad master equation (we set $\hbar=1$)
\begin{align}
\label{appendixA_proof_chapter1__MME}
        \frac{d}{d t}\hat{\rho} 
            &=-i[\hat{H},\; \hat{\rho}]
                +\sum\limits_{j=1}^{K}\mathfrak{D}[\hat{c}_{j}]\hat{\rho},
\end{align}
where $\hat{H}=\hat{H}^{\ast}$ is the system Hamiltonian, $\hat{c}=\begin{bmatrix}
\hat{c}_{1} &\hat{c}_{2} &\cdots &\hat{c}_{K}
\end{bmatrix}^{\top}$ is the vector of Lindblad operators, $K$ is the number of decoherence channels, and $\mathfrak{D}[\hat{c}_{j}]\hat{\rho} \triangleq \hat{c}_{j}\hat{\rho}\hat{c}_{j}^{\ast}-\frac{1}{2}\left(\hat{c}_{j}^{\ast}\hat{c}_{j}\hat{\rho}+ \hat{\rho} \hat{c}_{j}^{\ast}\hat{c}_{j} \right)$. 

Let $(\hat{q}_{j},\hat{p}_{j})$, $j=1,\cdots,N$, be the position and momentum operators of this quantum system. They satisfy the canonical commutation relation 
\begin{align}
\label{chapter1_commutation 1}
\left[\hat{x}, \hat{x}^{\top}\right]
=\hat{x}\hat{x}^{\top}-\left(\hat{x}\hat{x}^{\top}\right)^{\top}
  =i\Sigma, \quad \Sigma\triangleq\begin{bmatrix}
         0 & I_{N}\\
-I_{N} &0
\end{bmatrix}, 
\end{align}
where $\hat{x}\triangleq\begin{bmatrix}\hat{q}_{1} &\cdots &\hat{q}_{N} &\hat{p}_{1} &\cdots &\hat{p}_{N}\end{bmatrix}^{\top}$. Suppose that the system Hamiltonian $\hat{H}$ is quadratic in the quadrature operators; i.e., 
$\hat{H}=\frac{1}{2}\hat{x}^{\top}M\hat{x}$, with $M=M^{\top}\in \mathbb{R}^{2N \times 2N}$. Also, suppose that the vector of Lindblad operators $\hat{c}$ is  linear in the quadrature operators; i.e.,  $\hat{c} = C \hat{x}$, with $C\in \mathbb{C}^{K \times 2N}$. An open quantum system with such $\hat{H}$ and $\hat{c}$ is said to be a linear open quantum system. For a linear open quantum system, given Gaussian initial conditions, the Lindblad master equation~\eqref{appendixA_proof_chapter1__MME} will always have a Gaussian state as its solution.

The mean value of $\hat{x}$ is given by $\langle\hat{x}\rangle=\begin{bmatrix}\tr(\hat{x}_{1} \hat{\rho} ) &\tr(\hat{x}_{2} \hat{\rho} ) &\cdots & \tr(\hat{x}_{2N} \hat{\rho} )\end{bmatrix}^{\top}$ and the covariance matrix  of $\hat{x}$ is given by $V=\frac{1}{2}\langle \triangle\hat{x}{\triangle\hat{x}}^{\top}+(\triangle\hat{x}{\triangle\hat{x}}^{\top})^{\top} \rangle$, where $ \triangle\hat{x}=\hat{x}-\langle \hat{x} \rangle$. The moment evolution equations for $\langle \hat{x} \rangle$ and $V$ are given by 
\begin{numcases}{}
\frac{d\langle\hat{x}\rangle}{dt}=\mathcal{A}\langle\hat{x}\rangle, \label{chapter1_meanfunction} \\
\frac{dV}{dt}=\mathcal{A}V+V\mathcal{A}^{\top}+\mathcal{D},  \label{chapter1_covfunction}
\end{numcases}
where $\mathcal{A}=\Sigma\left(M+\im(C^{\dagger}C)\right)$ is the \emph{drift matrix} and 
$\mathcal{D}=\Sigma\re(C^{\dagger}C)\Sigma^{\top}$ is the \emph{diffusion matrix}. 

The result described in~\eqref{chapter1_meanfunction} and \eqref{chapter1_covfunction} and the explicit formulas for $\mathcal{A}$ and $\mathcal{D}$  can be found in several references; e.g., \cite{WD05:prl,DW05:QELSC,WM10:book,J14:book,GLS16:CP}. Recent results on the preparation of pure Gaussian states are all built upon the moment evolution equations \eqref{chapter1_meanfunction} and \eqref{chapter1_covfunction}~\cite{KY12:pra,Y12:ptrsa,IY13:pra,MWPY14:msc,NPF16:pra,MWPY18:auto,MPW17:scl,MWPY16:cdc2,MWPY17:jpa,M17:thesis}. To the best of the authors' knowledge, the explicit derivation for~\eqref{chapter1_meanfunction} and \eqref{chapter1_covfunction}  from the master equation~\eqref{appendixA_proof_chapter1__MME} cannot be found in the literature. So we provide a detailed derivation here for the interested reader.

\section{Derivation of Moment Evolution Equations}
For simplicity, we consider a linear open quantum system with a single decoherence channel ($K=1$). The extension of the derivation to multiple decoherence channels is straightforward; see Remark~\ref{remark}. We will first calculate the evolution equation for the mean vector $\langle\hat{x}\rangle$. Then, we will calculate the evolution equation for the covariance matrix $ V $. \\

\noindent\emph{\textbf{Part 1: Calculation of the Evolution Equation for the Mean Vector $\langle\hat{x}\rangle$}}

Let $\hat{x}_{\ell}$ be the $\ell$th entry of the column vector $\hat{x}$. Then the commutation relations~\eqref{chapter1_commutation 1} can be written as 
\begin{align*}
[\hat{x}_{\ell},\;\hat{x}_{m}]=\hat{x}_{\ell}\hat{x}_{m}-\hat{x}_{m}\hat{x}_{\ell}= i\Sigma_{\ell m}, 
\end{align*} 
where $\Sigma_{\ell m}$ is the $(\ell,m)$ entry of the matrix $\Sigma$. The equation of motion for $\langle\hat{x}_{\ell}\rangle$ is 
\begin{align}
\frac{d\langle\hat{x}_{\ell}\rangle}{dt}&=\tr\left(\hat{x}_{\ell}\frac{d \hat{\rho} }{d t}\right) = \tr\bigg(\hat{x}_{\ell}\left(-i[\hat{H},\; \hat{\rho}]+\mathfrak{D}[\hat{c}]\hat{\rho}\right)\bigg) \notag \\
&=\tr\bigg(\hat{x}_{\ell}\left(-i[\hat{H},\; \hat{\rho}]\right)\bigg)+\tr\bigg(\hat{x}_{\ell}\left(\mathfrak{D}[\hat{c}]\hat{\rho}\right)\bigg). \label{appendixA_proof_chapter1_meanfunction_evolution} 
\end{align}
Let us calculate $\tr\bigg(\hat{x}_{\ell}\left(-i[\hat{H},\; \hat{\rho}]\right)\bigg)$ and $\tr\bigg(\hat{x}_{\ell}\left(\mathfrak{D}[\hat{c}]\hat{\rho}\right)\bigg)$ separately. First, we have 
\begin{align}
&\tr\bigg(\hat{x}_{\ell}\left(-i[\hat{H},\hat{\rho}]\right)\bigg)=-\frac{i}{2}\tr\bigg(\hat{x}_{\ell}\hat{x}^{\top}M\hat{x}\hat{\rho}-\hat{x}_{\ell} \hat{\rho} \hat{x}^{\top}M\hat{x}  \bigg)\notag \\
=&-\frac{i}{2}\tr\bigg(\sum\limits_{j=1}^{2N}\sum\limits_{k=1}^{2N}M_{jk}\hat{x}_{\ell}\hat{x}_{j}\hat{x}_{k}\hat{\rho} - \sum\limits_{j=1}^{2N}\sum\limits_{k=1}^{2N}\hat{x}_{\ell} \hat{\rho} M_{jk}\hat{x}_{j}\hat{x}_{k}   \bigg)\notag \\
=&-\frac{i}{2}\sum\limits_{j=1}^{2N}\sum\limits_{k=1}^{2N} M_{jk} \tr \bigg(\hat{x}_{\ell}\hat{x}_{j}\hat{x}_{k}\hat{\rho} -  \hat{x}_{j}\hat{x}_{k} \hat{x}_{\ell}  \hat{\rho}\bigg) \notag \\
=&-\frac{i}{2}\sum\limits_{j=1}^{2N}\sum\limits_{k=1}^{2N}M_{jk}\tr\bigg(\left(\hat{x}_{j}\hat{x}_{\ell}+i\Sigma_{\ell j}\right)\hat{x}_{k}\hat{\rho} -  \hat{x}_{j}\hat{x}_{k} \hat{x}_{\ell}  \hat{\rho}\bigg)\notag \\
=&-\frac{i}{2}\sum\limits_{j=1}^{2N}\sum\limits_{k=1}^{2N}M_{jk} \tr\bigg(i\Sigma_{\ell j}\hat{x}_{k}\hat{\rho} +\hat{x}_{j}\hat{x}_{\ell}\hat{x}_{k}\hat{\rho} -  \hat{x}_{j}\hat{x}_{k} \hat{x}_{\ell}  \hat{\rho}\bigg)\notag \\
=&-\frac{i}{2}\sum\limits_{j=1}^{2N}\sum\limits_{k=1}^{2N}M_{jk}\tr\bigg(i\Sigma_{\ell j}\hat{x}_{k}\hat{\rho} +i\Sigma_{\ell k}\hat{x}_{j}\hat{\rho}\bigg)\notag \\
=&\frac{1}{2}\sum\limits_{j=1}^{2N}\sum\limits_{k=1}^{2N}\Sigma_{\ell j} M_{jk}\langle \hat{x}_{k}\rangle+
\frac{1}{2} \sum\limits_{j=1}^{2N}\sum\limits_{k=1}^{2N}\Sigma_{\ell k} M_{jk} \langle \hat{x}_{j}\rangle \notag \\
=&\frac{1}{2}\sum\limits_{j=1}^{2N}\sum\limits_{k=1}^{2N}\Sigma_{\ell j} M_{jk}\langle \hat{x}_{k}\rangle+
\frac{1}{2} \sum\limits_{j=1}^{2N}\sum\limits_{k=1}^{2N}\Sigma_{\ell k} M_{kj} \langle \hat{x}_{j}\rangle \notag \\
=&\sum\limits_{j=1}^{2N}\sum\limits_{k=1}^{2N}\Sigma_{\ell j} M_{jk}\langle \hat{x}_{k}\rangle\notag \\
=&\Sigma_{\ell :}M \langle \hat{x}\rangle, \label{appendixA_proof_chapter1_meanfunction_evolution1} 
\end{align}
where $\Sigma_{\ell :}$ denotes the $\ell$th row of $\Sigma$. Second, we have
\begin{align}
&\tr\bigg(\hat{x}_{\ell}\left(\mathfrak{D}[\hat{c}]\hat{\rho}\right)\bigg)=\tr\bigg(\hat{x}_{\ell}\Big(\hat{c}\hat{\rho}\hat{c}^{\ast}-\frac{1}{2} \hat{c}^{\ast}\hat{c}\hat{\rho}- \frac{1}{2}  \hat{\rho} \hat{c}^{\ast}\hat{c} \Big)\bigg)\notag \\
=&\tr\bigg(\hat{c}^{\ast}\hat{x}_{\ell} \hat{c}\hat{\rho}-\frac{1}{2} \hat{x}_{\ell} \hat{c}^{\ast}\hat{c}\hat{\rho}- \frac{1}{2}\hat{c}^{\ast}\hat{c}  \hat{x}_{\ell} \hat{\rho}  \bigg)\notag \\
=&\tr\bigg(\Big(\sum\limits_{j=1}^{2N} C_{j}^{\ast}\hat{x}_{j}\Big)\hat{x}_{\ell}\Big(\sum\limits_{k=1}^{2N} C_{k}\hat{x}_{k}\Big)\hat{\rho}-\frac{1}{2}\hat{x}_{\ell}\Big(\sum\limits_{j=1}^{2N} C_{j}^{\ast}\hat{x}_{j}\Big) \Big(\sum\limits_{k=1}^{2N} C_{k}\hat{x}_{k}\Big)\hat{\rho} \notag\\
&-\frac{1}{2} \Big(\sum\limits_{j=1}^{2N} C_{j}^{\ast}\hat{x}_{j}\Big)\Big(\sum\limits_{k=1}^{2N} C_{k}\hat{x}_{k}\Big) \hat{x}_{\ell} \hat{\rho}  \bigg)\notag \\
=&\sum\limits_{j=1}^{2N}\sum\limits_{k=1}^{2N}C_{j}^{\ast}C_{k} \tr\bigg(   \Big( \hat{x}_{j}\hat{x}_{\ell}\hat{x}_{k} -\frac{1}{2}\hat{x}_{\ell} \hat{x}_{j}\hat{x}_{k}-\frac{1}{2}   \hat{x}_{j} \hat{x}_{k} \hat{x}_{\ell} \Big)\hat{\rho}  \bigg)\notag \\
=&\sum\limits_{j=1}^{2N}\sum\limits_{k=1}^{2N} C_{j}^{\ast}C_{k}  \tr\bigg( \Big(\frac{i}{2}\Sigma_{j\ell}\hat{x}_{k}+\frac{i}{2}\Sigma_{\ell k}\hat{x}_{j}\Big)\hat{\rho}  \bigg)\notag \\
=&\frac{i}{2}\sum\limits_{j=1}^{2N}\sum\limits_{k=1}^{2N} C_{j}^{\ast}C_{k} \Sigma_{j\ell}\langle \hat{x}_{k}\rangle +\frac{i}{2} \sum\limits_{j=1}^{2N}\sum\limits_{k=1}^{2N} C_{j}^{\ast}C_{k}\Sigma_{\ell k}\langle \hat{x}_{j}\rangle\notag \\
=&- \frac{i}{2} \sum\limits_{j=1}^{2N}\sum\limits_{k=1}^{2N} \Sigma_{\ell j} C_{j}^{\ast}C_{k} \langle \hat{x}_{k}\rangle+\frac{i}{2}\sum\limits_{j=1}^{2N}\sum\limits_{k=1}^{2N} \Sigma_{\ell k}C_{k} C_{j}^{\ast} \langle \hat{x}_{j}\rangle\notag \\
=&- \frac{i}{2} \sum\limits_{j=1}^{2N}\sum\limits_{k=1}^{2N} \Sigma_{\ell j} C_{j}^{\ast}C_{k} \langle \hat{x}_{k}\rangle+\frac{i}{2}\sum\limits_{j=1}^{2N}\sum\limits_{k=1}^{2N} \Sigma_{\ell j}C_{j} C_{k}^{\ast} \langle \hat{x}_{k}\rangle\notag \\
=&\frac{i}{2}\sum\limits_{j=1}^{2N}\sum\limits_{k=1}^{2N}\Sigma_{\ell j}\left( - C_{j}^{\ast}C_{k}+ C_{j}C_{k}^{\ast} \right) \langle \hat{x}_{k}\rangle\notag \\
=&\sum\limits_{j=1}^{2N}\sum\limits_{k=1}^{2N}\Sigma_{\ell j} \im \left( C_{j}^{\ast}C_{k} \right) \langle \hat{x}_{k}\rangle \notag \\
=&  \Sigma_{\ell :}\im\left(C^{\dagger}C \right) \langle \hat{x}\rangle.  \label{appendixA_proof_chapter1_meanfunction_evolution2} 
\end{align}
Here $C_{j}\in\mathbb{C}$ denotes the $j$th entry of the row vector $C$. Substituting~\eqref{appendixA_proof_chapter1_meanfunction_evolution1} and~\eqref{appendixA_proof_chapter1_meanfunction_evolution2} into~\eqref{appendixA_proof_chapter1_meanfunction_evolution}, we obtain 
\begin{align}
\frac{d\langle\hat{x}_{\ell}\rangle}{dt}&=\Sigma_{\ell :}\left(M + \im\left(C^{\dagger}C \right) \right) \langle \hat{x}\rangle. \label{appendixA_proof_chapter1_meanfunction_evolution3}
\end{align}
The evolution equation \eqref{chapter1_meanfunction} follows immediately from \eqref{appendixA_proof_chapter1_meanfunction_evolution3}. \\\vspace{0.4cm}

\noindent\emph{\textbf{Part 2: Calculation of the Evolution Equation for the Covariance Matrix $V$}} \vspace{0.4cm}

Suppose $V_{\ell m}$ is the $(\ell,m)$ entry of the covariance matrix $V$. Then we have 
\begin{align*}
&V_{\ell m}\notag\\
=&\frac{1}{2}\langle\triangle\hat{x}_{\ell}\triangle\hat{x}_{m}+\triangle\hat{x}_{m}\triangle\hat{x}_{\ell}\rangle \\
=&\frac{1}{2}\langle(\hat{x}_{\ell}-\langle\hat{x}_{\ell}\rangle)(\hat{x}_{m}-\langle\hat{x}_{m}\rangle)+(\hat{x}_{m}-\langle\hat{x}_{m}\rangle)(\hat{x}_{\ell}-\langle\hat{x}_{\ell}\rangle)\rangle \\
=&\frac{1}{2}\left(\langle\hat{x}_{\ell}\hat{x}_{m}+ \hat{x}_{m}\hat{x}_{\ell} \rangle -\langle\hat{x}_{\ell}\rangle\langle\hat{x}_{m}\rangle-\langle\hat{x}_{m}\rangle\langle\hat{x}_{\ell}\rangle  \right)\\
=&\frac{1}{2}\langle\hat{x}_{\ell}\hat{x}_{m}+ \hat{x}_{m}\hat{x}_{\ell} \rangle -\langle\hat{x}_{\ell}\rangle\langle\hat{x}_{m}\rangle.
\end{align*}

Therefore, the equation of motion for $V_{\ell m}$ is given by 
\begin{align}
&\frac{dV_{\ell m}}{dt}\notag\\
=&\frac{1}{2}\frac{d\langle\hat{x}_{\ell}\hat{x}_{m}+ \hat{x}_{m}\hat{x}_{\ell} \rangle}{dt} -\frac{d\left(\langle\hat{x}_{\ell}\rangle\langle\hat{x}_{m}\rangle\right)}{dt}\notag\\
=&\frac{1}{2} \tr\bigg(\left(\hat{x}_{\ell}\hat{x}_{m}+\hat{x}_{m}\hat{x}_{\ell}\right)\frac{d\hat{\rho}}{dt}\bigg)-\frac{d\left(\langle\hat{x}_{\ell}\rangle\langle\hat{x}_{m}\rangle\right)}{dt}\notag\\ 
=&\frac{1}{2}\tr\bigg(\left(\hat{x}_{\ell}\hat{x}_{m}+\hat{x}_{m}\hat{x}_{\ell}\right)\left(-i[\hat{H},\hat{\rho}]\right)\bigg)\notag\\
&+ \frac{1}{2}\tr\bigg(\left(\hat{x}_{\ell}\hat{x}_{m}+\hat{x}_{m}\hat{x}_{\ell}\right)\left(\mathfrak{D}[\hat{c}]\hat{\rho}\right)\bigg)  -\frac{d\left(\langle\hat{x}_{\ell}\rangle\langle\hat{x}_{m}\rangle\right)}{dt}.\label{appendixA_proof_chapter1_covariance_evolution} 
\end{align}

Let us calculate $\tr\bigg(\left(\hat{x}_{\ell}\hat{x}_{m}+\hat{x}_{m}\hat{x}_{\ell}\right)\left(-i[\hat{H},\hat{\rho}]\right)\bigg)$, $\tr\bigg(\left(\hat{x}_{\ell}\hat{x}_{m}+\hat{x}_{m}\hat{x}_{\ell}\right)\left(\mathfrak{D}[\hat{c}]\hat{\rho}\right)\bigg)$ and $\frac{d\left(\langle\hat{x}_{\ell}\rangle\langle\hat{x}_{m}\rangle\right)}{dt}$ separately. First, we have 
\begin{align}
&\tr\bigg(\left(\hat{x}_{\ell}\hat{x}_{m}+\hat{x}_{m}\hat{x}_{\ell}\right)\left(-i[\hat{H},\hat{\rho}]\right)\bigg)\notag \\
=&-\frac{i}{2}\tr\bigg(\sum\limits_{j=1}^{2N}\sum\limits_{k=1}^{2N}\left(\hat{x}_{\ell}\hat{x}_{m}+\hat{x}_{m}\hat{x}_{\ell}\right)M_{jk}\hat{x}_{j}\hat{x}_{k}\hat{\rho} \notag\\
&- \sum\limits_{j=1}^{2N}\sum\limits_{k=1}^{2N}\left(\hat{x}_{\ell}\hat{x}_{m}+\hat{x}_{m}\hat{x}_{\ell}\right) \hat{\rho} M_{jk}\hat{x}_{j}\hat{x}_{k} \bigg)\notag \\
=&-\frac{i}{2}\sum\limits_{j=1}^{2N}\sum\limits_{k=1}^{2N}M_{jk} \tr\bigg(\hat{x}_{\ell}\hat{x}_{m}\hat{x}_{j}\hat{x}_{k}\hat{\rho}+\hat{x}_{m}\hat{x}_{\ell}\hat{x}_{j}\hat{x}_{k}\hat{\rho} \notag \\
& - \hat{x}_{\ell}\hat{x}_{m} \hat{\rho} \hat{x}_{j}\hat{x}_{k}  - \hat{x}_{m}\hat{x}_{\ell} \hat{\rho} \hat{x}_{j}\hat{x}_{k}   \bigg) \notag \\
=&-\frac{i}{2}\sum\limits_{j=1}^{2N}\sum\limits_{k=1}^{2N}M_{jk}\tr \bigg(\hat{x}_{\ell}\hat{x}_{m}\hat{x}_{j}\hat{x}_{k}\hat{\rho}+\hat{x}_{m}\hat{x}_{\ell}\hat{x}_{j}\hat{x}_{k}\hat{\rho} \notag \\
& - \hat{x}_{j}\hat{x}_{k} \hat{x}_{\ell}\hat{x}_{m} \hat{\rho}  -  \hat{x}_{j}\hat{x}_{k}  \hat{x}_{m}\hat{x}_{\ell} \hat{\rho}  \bigg)  \notag \\
=&-\frac{i}{2}\sum\limits_{j=1}^{2N}\sum\limits_{k=1}^{2N}M_{jk} \tr\bigg(\hat{x}_{\ell}\hat{x}_{m}\hat{x}_{j}\hat{x}_{k}\hat{\rho}+\left(\hat{x}_{\ell}\hat{x}_{m}+i\Sigma_{m \ell}\right)\hat{x}_{j}\hat{x}_{k}\hat{\rho}\notag \\
& - \hat{x}_{j}\hat{x}_{k} \hat{x}_{\ell}\hat{x}_{m} \hat{\rho}   -\hat{x}_{j}\hat{x}_{k}\left(\hat{x}_{\ell}\hat{x}_{m}+i\Sigma_{m \ell}\right) \hat{\rho}    \bigg) \notag \\
=&-i\sum\limits_{j=1}^{2N}\sum\limits_{k=1}^{2N}M_{jk} \tr \bigg(\hat{x}_{\ell}\hat{x}_{m}\hat{x}_{j}\hat{x}_{k}\hat{\rho}- \hat{x}_{j}\hat{x}_{k} \hat{x}_{\ell}\hat{x}_{m} \hat{\rho}   \bigg) \notag \\
=&-i\sum\limits_{j=1}^{2N}\sum\limits_{k=1}^{2N}M_{jk}\tr \bigg(\hat{x}_{\ell}\left(\hat{x}_{j}\hat{x}_{m}+i\Sigma_{mj}\right)\hat{x}_{k}\hat{\rho}- \hat{x}_{j}\left(\hat{x}_{\ell}\hat{x}_{k}+i\Sigma_{k\ell}\right) \hat{x}_{m} \hat{\rho}    \bigg)\notag \\
=&-i\sum\limits_{j=1}^{2N}\sum\limits_{k=1}^{2N}M_{jk} \tr \bigg(\hat{x}_{\ell}\hat{x}_{j}\hat{x}_{m}\hat{x}_{k}\hat{\rho}- \hat{x}_{j}\hat{x}_{\ell}\hat{x}_{k} \hat{x}_{m} \hat{\rho} \notag\\
& +i\Sigma_{mj} \hat{x}_{\ell}\hat{x}_{k} \hat{\rho} -i\Sigma_{k\ell} \hat{x}_{j} \hat{x}_{m} \hat{\rho}
\bigg)  \notag \\
=&-i\sum\limits_{j=1}^{2N}\sum\limits_{k=1}^{2N}M_{jk} \tr \bigg(\hat{x}_{\ell}\hat{x}_{j}\left(\hat{x}_{k}\hat{x}_{m}+i\Sigma_{mk}\right)\hat{\rho}- \left( \hat{x}_{\ell}\hat{x}_{j} +i\Sigma_{j\ell} \right) \hat{x}_{k} \hat{x}_{m} \hat{\rho} \notag \\
& +i\Sigma_{mj} \hat{x}_{\ell}\hat{x}_{k} \hat{\rho} -i\Sigma_{k\ell} \hat{x}_{j} \hat{x}_{m} \hat{\rho}
\bigg) \notag \\
=&-i\sum\limits_{j=1}^{2N}\sum\limits_{k=1}^{2N}M_{jk} \tr \bigg(i\Sigma_{mk} \hat{x}_{\ell}\hat{x}_{j}\hat{\rho}- i\Sigma_{j\ell}    \hat{x}_{k} \hat{x}_{m} \hat{\rho} \notag\\
& +i\Sigma_{mj} \hat{x}_{\ell}\hat{x}_{k} \hat{\rho} -i\Sigma_{k\ell} \hat{x}_{j} \hat{x}_{m} \hat{\rho}\bigg)  \notag \\
=&\sum\limits_{j=1}^{2N}\sum\limits_{k=1}^{2N}M_{jk}\Sigma_{mk} \tr \bigg( \hat{x}_{\ell}\hat{x}_{j}\hat{\rho}\bigg) -\sum\limits_{j=1}^{2N}\sum\limits_{k=1}^{2N}M_{jk}\Sigma_{j\ell} \tr \bigg(  \hat{x}_{k} \hat{x}_{m} \hat{\rho}\bigg) \notag\notag \\
& +\sum\limits_{j=1}^{2N}\sum\limits_{k=1}^{2N}M_{jk}\Sigma_{mj} \tr \bigg( \hat{x}_{\ell}\hat{x}_{k}\hat{\rho}\bigg)- \sum\limits_{j=1}^{2N}\sum\limits_{k=1}^{2N}M_{jk}\Sigma_{k\ell} \tr \bigg( \hat{x}_{j} \hat{x}_{m} \hat{\rho}\bigg) \notag \\
=&2\sum\limits_{j=1}^{2N}\sum\limits_{k=1}^{2N}M_{jk}\Sigma_{mk} \tr \bigg( \hat{x}_{\ell}\hat{x}_{j}\hat{\rho}\bigg) -2\sum\limits_{j=1}^{2N}\sum\limits_{k=1}^{2N}M_{jk}\Sigma_{j\ell}  \tr \bigg(  \hat{x}_{k} \hat{x}_{m} \hat{\rho}\bigg) \notag \\
=&\sum\limits_{j=1}^{2N}\sum\limits_{k=1}^{2N}M_{jk}\Sigma_{mk} \tr\bigg(\left(\hat{x}_{\ell}\hat{x}_{j}+ \hat{x}_{j}\hat{x}_{\ell}+i\Sigma_{\ell j} \right)\hat{\rho} \bigg)\notag \\
& -\sum\limits_{j=1}^{2N}\sum\limits_{k=1}^{2N}M_{jk}\Sigma_{j\ell}  \tr\bigg( \left(    \hat{x}_{k} \hat{x}_{m} +  \hat{x}_{m}\hat{x}_{k}+i\Sigma_{km}\right) \hat{\rho} \bigg)\notag \\
=&\sum\limits_{j=1}^{2N}\sum\limits_{k=1}^{2N}M_{jk}\Sigma_{mk}  \langle \hat{x}_{\ell}\hat{x}_{j}+ \hat{x}_{j}\hat{x}_{\ell}\rangle - \sum\limits_{j=1}^{2N}\sum\limits_{k=1}^{2N}M_{jk}\Sigma_{j\ell}  \langle    \hat{x}_{k} \hat{x}_{m} +  \hat{x}_{m}\hat{x}_{k}\rangle \notag \\
=&\sum\limits_{j=1}^{2N}\sum\limits_{k=1}^{2N}\Sigma_{\ell j} M_{jk}  \langle  \hat{x}_{k} \hat{x}_{m} +  \hat{x}_{m}\hat{x}_{k}\rangle - \sum\limits_{j=1}^{2N}\sum\limits_{k=1}^{2N} \langle \hat{x}_{\ell}\hat{x}_{j}+ \hat{x}_{j}\hat{x}_{\ell}\rangle M_{jk} \Sigma_{km}. \label{appendixA_proof_chapter1_covariance_evolution1} 
\end{align} \vspace{0.4cm}

Second, we have 
\begin{align}
&\tr\bigg(\left(\hat{x}_{\ell}\hat{x}_{m}+\hat{x}_{m}\hat{x}_{\ell}\right)\left(\mathfrak{D}[\hat{c}]\hat{\rho}\right)\bigg)\notag \\
=&\tr\bigg(\left(\hat{x}_{\ell}\hat{x}_{m}+\hat{x}_{m}\hat{x}_{\ell}\right)\Big(\hat{c}\hat{\rho}\hat{c}^{\ast}-\frac{1}{2}\hat{c}^{\ast}\hat{c}\hat{\rho}-\frac{1}{2} \hat{\rho} \hat{c}^{\ast}\hat{c}\Big)\bigg)\notag \\
=&\tr\bigg(\hat{c}^{\ast} \left(\hat{x}_{\ell}\hat{x}_{m}+\hat{x}_{m}\hat{x}_{\ell}\right)\hat{c}\hat{\rho}-\frac{1}{2}\left(\hat{x}_{\ell}\hat{x}_{m}+\hat{x}_{m}\hat{x}_{\ell}\right)\hat{c}^{\ast}\hat{c}\hat{\rho} \notag\\
&-\frac{1}{2}\hat{c}^{\ast}\hat{c} \left(\hat{x}_{\ell}\hat{x}_{m}+\hat{x}_{m}\hat{x}_{\ell}\right) \hat{\rho} \bigg)\notag \\
=&\tr\bigg(\Big(\sum\limits_{j=1}^{2N} C_{j}^{\ast}\hat{x}_{j}\Big)\Big(\hat{x}_{\ell}\hat{x}_{m}+\hat{x}_{m}\hat{x}_{\ell}\Big)\Big(\sum\limits_{k=1}^{2N} C_{k}\hat{x}_{k}\Big)\hat{\rho}\notag \\
& -\frac{1}{2}\Big(\hat{x}_{\ell}\hat{x}_{m}+\hat{x}_{m}\hat{x}_{\ell}\Big)\Big(\sum\limits_{j=1}^{2N} C_{j}^{\ast}\hat{x}_{j}\Big) \Big(\sum\limits_{k=1}^{2N} C_{k}\hat{x}_{k}\Big)\hat{\rho}\notag \\
& -\frac{1}{2} \Big(\sum\limits_{j=1}^{2N} C_{j}^{\ast}\hat{x}_{j}\Big)\Big(\sum\limits_{k=1}^{2N} C_{k}\hat{x}_{k}\Big)\Big(\hat{x}_{\ell}\hat{x}_{m}+\hat{x}_{m}\hat{x}_{\ell}\Big) \hat{\rho}  \bigg)\notag \\
=&\sum\limits_{j=1}^{2N}\sum\limits_{k=1}^{2N} C_{j}^{\ast}C_{k} \tr\bigg( \Big( \hat{x}_{j}\left(\hat{x}_{\ell}\hat{x}_{m}+\hat{x}_{m}\hat{x}_{\ell}\right)\hat{x}_{k} \notag \\
&-\frac{1}{2}\left(\hat{x}_{\ell}\hat{x}_{m}+\hat{x}_{m}\hat{x}_{\ell}\right) \hat{x}_{j}\hat{x}_{k}
-\frac{1}{2}   \hat{x}_{j} \hat{x}_{k} \left(\hat{x}_{\ell}\hat{x}_{m}+\hat{x}_{m}\hat{x}_{\ell}\right) \Big)\hat{\rho}  \bigg)\notag \\
=&\sum\limits_{j=1}^{2N}\sum\limits_{k=1}^{2N} C_{j}^{\ast}C_{k} \tr\bigg(  \Big( \hat{x}_{j}\left(2\hat{x}_{\ell}\hat{x}_{m}-i\Sigma_{\ell m}\right)\hat{x}_{k} \notag \\
& - \frac{1}{2} \left(2\hat{x}_{\ell}\hat{x}_{m}-i\Sigma_{\ell m}\right)  \hat{x}_{j}\hat{x}_{k} -  \frac{1}{2}   \hat{x}_{j} \hat{x}_{k} \left(2\hat{x}_{\ell}\hat{x}_{m}-i\Sigma_{\ell m}\right)\Big)\hat{\rho}  \bigg)\notag \\
=&\sum\limits_{j=1}^{2N}\sum\limits_{k=1}^{2N} C_{j}^{\ast}C_{k} \tr\bigg(  \Big( 2 \hat{x}_{j} \hat{x}_{\ell}\hat{x}_{m} \hat{x}_{k}  -  \hat{x}_{\ell}\hat{x}_{m} \hat{x}_{j}\hat{x}_{k} -   \hat{x}_{j} \hat{x}_{k}  \hat{x}_{\ell}\hat{x}_{m} \Big)\hat{\rho}  \bigg)\notag \\
=&\sum\limits_{j=1}^{2N}\sum\limits_{k=1}^{2N} C_{j}^{\ast}C_{k} \tr\bigg( \Big( 2 \left(\hat{x}_{\ell} \hat{x}_{j} +i\Sigma_{j\ell}\right) \left(\hat{x}_{k} \hat{x}_{m}+i\Sigma_{mk} \right)  \notag\\ &-  \hat{x}_{\ell}\left( \hat{x}_{j} \hat{x}_{m}+i\Sigma_{mj}\right)\hat{x}_{k} -   \hat{x}_{j} \left( \hat{x}_{\ell} \hat{x}_{k}+i\Sigma_{k\ell} \right)\hat{x}_{m} \Big)\hat{\rho} \bigg)\notag \\
=&\sum\limits_{j=1}^{2N}\sum\limits_{k=1}^{2N} C_{j}^{\ast}C_{k} \tr\bigg( \Big(  2\hat{x}_{\ell} \hat{x}_{j}\hat{x}_{k} \hat{x}_{m} +2i\Sigma_{j\ell}\hat{x}_{k} \hat{x}_{m}+2i\Sigma_{mk} \hat{x}_{\ell} \hat{x}_{j} \notag \\
& - 2\Sigma_{j\ell}\Sigma_{mk}  -  \hat{x}_{\ell} \hat{x}_{j} \hat{x}_{m}\hat{x}_{k}-i\Sigma_{mj}  \hat{x}_{\ell} \hat{x}_{k} -   \hat{x}_{j}  \hat{x}_{\ell} \hat{x}_{k}\hat{x}_{m}  - i\Sigma_{k\ell}  \hat{x}_{j}  \hat{x}_{m} \Big)\hat{\rho}  \bigg)\notag \\
=&\sum\limits_{j=1}^{2N}\sum\limits_{k=1}^{2N} C_{j}^{\ast}C_{k}   \tr\bigg(\Big( i\Sigma_{km}\hat{x}_{\ell} \hat{x}_{j} + i\Sigma_{\ell j}\hat{x}_{k} \hat{x}_{m}+2 i\Sigma_{j\ell}\hat{x}_{k} \hat{x}_{m}\notag\\
&+2 i\Sigma_{mk} \hat{x}_{\ell} \hat{x}_{j} - 2 \Sigma_{j\ell}\Sigma_{mk}  -i\Sigma_{mj}  \hat{x}_{\ell} \hat{x}_{k}   - i\Sigma_{k\ell}  \hat{x}_{j}  \hat{x}_{m} \Big)\hat{\rho}  \bigg)\notag \\
=&\sum\limits_{j=1}^{2N}\sum\limits_{k=1}^{2N} C_{j}^{\ast}C_{k}  \tr\bigg(\Big(  i\Sigma_{j\ell}\hat{x}_{k} \hat{x}_{m}+ i\Sigma_{mk} \hat{x}_{\ell} \hat{x}_{j} \notag \\
&- 2 \Sigma_{j\ell}\Sigma_{mk} -i\Sigma_{mj}  \hat{x}_{\ell} \hat{x}_{k}   - i\Sigma_{k\ell}  \hat{x}_{j}  \hat{x}_{m} \Big)\hat{\rho}  \bigg)\notag \\
=&\sum\limits_{j=1}^{2N}\sum\limits_{k=1}^{2N}i C_{j}^{\ast}C_{k}\Sigma_{j\ell} \tr\bigg(  \hat{x}_{k} \hat{x}_{m}\hat{\rho}\bigg) + \sum\limits_{j=1}^{2N}\sum\limits_{k=1}^{2N} i C_{j}^{\ast}C_{k} \Sigma_{mk}  \tr\bigg(  \hat{x}_{\ell} \hat{x}_{j} \hat{\rho} \bigg) \notag\\
&- 2  \sum\limits_{j=1}^{2N}\sum\limits_{k=1}^{2N} C_{j}^{\ast}C_{k}   \Sigma_{j\ell}\Sigma_{mk}  - \sum\limits_{j=1}^{2N}\sum\limits_{k=1}^{2N}i C_{j}^{\ast}C_{k}\Sigma_{mj} \tr\bigg(     \hat{x}_{\ell} \hat{x}_{k}\hat{\rho} \bigg)  \notag\\
& - \sum\limits_{j=1}^{2N}\sum\limits_{k=1}^{2N} i C_{j}^{\ast}C_{k}\Sigma_{k\ell} \tr\bigg(  \hat{x}_{j}  \hat{x}_{m} \hat{\rho}  \bigg)\notag \\
=&\sum\limits_{j=1}^{2N}\sum\limits_{k=1}^{2N}i C_{j}^{\ast}C_{k} \Sigma_{j\ell} \tr\bigg( \hat{x}_{k} \hat{x}_{m}\hat{\rho}\bigg)  + \sum\limits_{j=1}^{2N}\sum\limits_{k=1}^{2N}i C_{j}^{\ast}C_{k}\Sigma_{mk} \tr\bigg(   \hat{x}_{\ell} \hat{x}_{j} \hat{\rho} \bigg) \notag\\
&- 2 \sum\limits_{j=1}^{2N}\sum\limits_{k=1}^{2N} C_{j}^{\ast}C_{k}   \Sigma_{j\ell}\Sigma_{mk} - \sum\limits_{j=1}^{2N}\sum\limits_{k=1}^{2N}i C_{k}^{\ast} C_{j}  \Sigma_{mk}  \tr\bigg(   \hat{x}_{\ell} \hat{x}_{j}\hat{\rho} \bigg)  \notag\\
&  - \sum\limits_{j=1}^{2N}\sum\limits_{k=1}^{2N}i C_{k}^{\ast} C_{j} \Sigma_{j\ell}  \tr\bigg(   \hat{x}_{k}  \hat{x}_{m} \hat{\rho} \bigg) \notag \\
=&\sum\limits_{j=1}^{2N}\sum\limits_{k=1}^{2N}i \left( C_{j}^{\ast}C_{k}- C_{k}^{\ast} C_{j}  \right)\Sigma_{j\ell} \tr\bigg(\hat{x}_{k} \hat{x}_{m}\hat{\rho}\bigg) \notag\\
&+ \sum\limits_{j=1}^{2N}\sum\limits_{k=1}^{2N}i\left( C_{j}^{\ast}C_{k}- C_{k}^{\ast}  C_{j}  \right)\Sigma_{mk} \tr\bigg(  \hat{x}_{\ell} \hat{x}_{j} \hat{\rho} \bigg) \notag \\
& - 2 \sum\limits_{j=1}^{2N}\sum\limits_{k=1}^{2N} C_{j}^{\ast}C_{k}   \Sigma_{j\ell}\Sigma_{mk}\notag \\
=&\sum\limits_{j=1}^{2N}\sum\limits_{k=1}^{2N}2\im\left(  C_{j} C_{k}^{\ast} \right) \Sigma_{j\ell} \tr\bigg(\hat{x}_{k} \hat{x}_{m}\hat{\rho}\bigg) \notag \\
&+ \sum\limits_{j=1}^{2N}\sum\limits_{k=1}^{2N}2\im\left( C_{j} C_{k}^{\ast}  \right) \Sigma_{mk} \tr\bigg(   \hat{x}_{\ell} \hat{x}_{j} \hat{\rho} \bigg)\notag \\
& - 2 \sum\limits_{j=1}^{2N}\sum\limits_{k=1}^{2N} C_{j}^{\ast}C_{k}   \Sigma_{j\ell}\Sigma_{mk} \notag \\
=&\sum\limits_{j=1}^{2N}\sum\limits_{k=1}^{2N}2\im\left(  C_{j} C_{k}^{\ast} \right)\Sigma_{j\ell} \tr\bigg(  \frac{\hat{x}_{k} \hat{x}_{m}+\hat{x}_{m}\hat{x}_{k}+i\Sigma_{km} }{2}\hat{\rho} \bigg)\notag \\
& + \sum\limits_{j=1}^{2N}\sum\limits_{k=1}^{2N}2\im\left( C_{j} C_{k}^{\ast}  \right)\Sigma_{mk} \tr\bigg( \frac{\hat{x}_{\ell} \hat{x}_{j}+\hat{x}_{j}\hat{x}_{\ell}+i\Sigma_{\ell j} }{2} \hat{\rho} \bigg) \notag\\
& - 2 \sum\limits_{j=1}^{2N}\sum\limits_{k=1}^{2N} C_{j}^{\ast}C_{k}   \Sigma_{j\ell}\Sigma_{mk} \notag \\
=&\sum\limits_{j=1}^{2N}\sum\limits_{k=1}^{2N}\im\left(  C_{j} C_{k}^{\ast} \right)  \Sigma_{j\ell}\langle\hat{x}_{k} \hat{x}_{m}+\hat{x}_{m}\hat{x}_{k}\rangle \notag\\
&+ \sum\limits_{j=1}^{2N}\sum\limits_{k=1}^{2N}\im\left(  C_{j} C_{k}^{\ast} \right)\left( i \Sigma_{j\ell} \Sigma_{km} \right)  \notag \\
&  + \sum\limits_{j=1}^{2N}\sum\limits_{k=1}^{2N}\im\left( C_{j} C_{k}^{\ast}  \right) \Sigma_{mk} \langle \hat{x}_{\ell} \hat{x}_{j}+\hat{x}_{j}\hat{x}_{\ell} \rangle \notag\\
& + \sum\limits_{j=1}^{2N}\sum\limits_{k=1}^{2N}\im\left( C_{j} C_{k}^{\ast}  \right) \left(  i \Sigma_{mk} \Sigma_{\ell j} \right) \notag \\
&  - 2 \sum\limits_{j=1}^{2N}\sum\limits_{k=1}^{2N} C_{j}^{\ast}C_{k}   \Sigma_{j\ell}\Sigma_{mk} \notag \\
=&\sum\limits_{j=1}^{2N}\sum\limits_{k=1}^{2N}\Sigma_{\ell j}\im\left(  C_{j}^{\ast} C_{k} \right)  \langle \hat{x}_{k} \hat{x}_{m}+\hat{x}_{m}\hat{x}_{k}\rangle \notag\\
&+2i \sum\limits_{j=1}^{2N}\sum\limits_{k=1}^{2N} \Sigma_{\ell j} \im\left(  C_{j}^{\ast} C_{k} \right)  \Sigma_{km}  \notag \\
&  +  \sum\limits_{j=1}^{2N}\sum\limits_{k=1}^{2N} \langle \hat{x}_{\ell} \hat{x}_{j}+\hat{x}_{j}\hat{x}_{\ell} \rangle \im\left( C_{j}^{\ast}  C_{k} \right) \Sigma_{k m}   \notag\\
& - 2 \sum\limits_{j=1}^{2N}\sum\limits_{k=1}^{2N}  \Sigma_{\ell j} C_{j}^{\ast}C_{k}  \Sigma_{k m} \notag \\
=&\sum\limits_{j=1}^{2N}\sum\limits_{k=1}^{2N}\Sigma_{\ell j}\im\left(  C_{j}^{\ast} C_{k} \right)  \langle \hat{x}_{k} \hat{x}_{m}+\hat{x}_{m}\hat{x}_{k}\rangle \notag\\
&+ \sum\limits_{j=1}^{2N}\sum\limits_{k=1}^{2N} \langle \hat{x}_{\ell} \hat{x}_{j}+\hat{x}_{j}\hat{x}_{\ell} \rangle \im\left( C_{j}^{\ast}  C_{k} \right) \Sigma_{k m} \notag \\
& - 2 \sum\limits_{j=1}^{2N}\sum\limits_{k=1}^{2N}  \Sigma_{\ell j} \re\left( C_{j}^{\ast}C_{k} \right) \Sigma_{k m}.  \label{appendixA_proof_chapter1_covariance_evolution2} 
\end{align}

Third, using~\eqref{appendixA_proof_chapter1_meanfunction_evolution3}, we have 
\begin{align}
&\frac{d\langle\hat{x}_{\ell}\rangle\langle\hat{x}_{m}\rangle}{dt}=\frac{d\langle\hat{x}_{\ell}\rangle}{dt}\langle\hat{x}_{m}\rangle+ \langle\hat{x}_{\ell}\rangle \frac{d\langle\hat{x}_{m}\rangle}{dt}\notag \\
=&\bigg(\sum\limits_{j=1}^{2N}\sum\limits_{k=1}^{2N}\Sigma_{\ell j} M_{jk}\langle \hat{x}_{k}\rangle + \sum\limits_{j=1}^{2N}\sum\limits_{k=1}^{2N}\Sigma_{\ell j} \im \left( C_{j}^{\ast}C_{k} \right) \langle \hat{x}_{k}\rangle\bigg)  \langle\hat{x}_{m}\rangle  \notag \\
&+\langle\hat{x}_{\ell}\rangle \bigg(\sum\limits_{j=1}^{2N}\sum\limits_{k=1}^{2N}\Sigma_{m j} M_{jk}\langle \hat{x}_{k}\rangle + \sum\limits_{j=1}^{2N}\sum\limits_{k=1}^{2N}\Sigma_{m j} \im \left( C_{j}^{\ast}C_{k} \right) \langle \hat{x}_{k}\rangle\bigg)  \notag\\
=&\bigg(\sum\limits_{j=1}^{2N}\sum\limits_{k=1}^{2N}\Sigma_{\ell j} M_{jk}\langle \hat{x}_{k}\rangle + \sum\limits_{j=1}^{2N}\sum\limits_{k=1}^{2N}\Sigma_{\ell j} \im \left( C_{j}^{\ast}C_{k} \right) \langle \hat{x}_{k}\rangle\bigg)  \langle\hat{x}_{m}\rangle  \notag \\
+&\langle\hat{x}_{\ell}\rangle \left(-\sum\limits_{j=1}^{2N}\sum\limits_{k=1}^{2N}\langle\hat{x}_{k}\rangle  M_{kj}   \Sigma_{j m }+  \sum\limits_{j=1}^{2N}\sum\limits_{k=1}^{2N}  \langle \hat{x}_{k}\rangle \im \left( C_{k}^{\ast}C_{j} \right) \Sigma_{jm}\right) \notag\\
=&\sum\limits_{j=1}^{2N}\sum\limits_{k=1}^{2N}\Sigma_{\ell j} \bigg( M_{jk} + \im \left( C_{j}^{\ast}C_{k} \right) \bigg) \langle \hat{x}_{k} \rangle \langle\hat{x}_{m}\rangle  \notag \\
&+ \sum\limits_{j=1}^{2N}\sum\limits_{k=1}^{2N} \langle\hat{x}_{\ell}\rangle \langle\hat{x}_{k}\rangle \bigg(  -M_{kj}   + \im \left( C_{k}^{\ast}C_{j} \right)\bigg) \Sigma_{jm} \notag\\
=&\sum\limits_{j=1}^{2N}\sum\limits_{k=1}^{2N}\Sigma_{\ell j} \bigg( M_{jk} + \im \left( C_{j}^{\ast}C_{k} \right) \bigg) \langle \hat{x}_{k} \rangle \langle\hat{x}_{m}\rangle  \notag \\
&+ \sum\limits_{j=1}^{2N}\sum\limits_{k=1}^{2N} \langle\hat{x}_{\ell}\rangle \langle\hat{x}_{j}\rangle \bigg(  -M_{jk}   + \im \left( C_{j}^{\ast}C_{k} \right)\bigg) \Sigma_{km}.    \label{appendixA_proof_chapter1_covariance_evolution3}  
\end{align}

Substituting~\eqref{appendixA_proof_chapter1_covariance_evolution1}, \eqref{appendixA_proof_chapter1_covariance_evolution2}  and~\eqref{appendixA_proof_chapter1_covariance_evolution3} into~\eqref{appendixA_proof_chapter1_covariance_evolution}, we obtain 
\begin{align}
&\frac{dV_{\ell m}}{dt} \notag\\
=&\frac{1}{2}\tr\bigg(\left(\hat{x}_{\ell}\hat{x}_{m}+\hat{x}_{m}\hat{x}_{\ell}\right)\left(-i[\hat{H},\hat{\rho}]\right)\bigg) \notag\\
&+ \frac{1}{2}\tr\bigg(\left(\hat{x}_{\ell}\hat{x}_{m}+\hat{x}_{m}\hat{x}_{\ell}\right)\left(\mathfrak{D}[\hat{c}]\hat{\rho}\right)\bigg)  -\frac{d\left(\langle\hat{x}_{\ell}\rangle\langle\hat{x}_{m}\rangle\right)}{dt} \notag \\
=&\sum\limits_{j=1}^{2N}\sum\limits_{k=1}^{2N}\Sigma_{\ell j} M_{jk} \frac{  \langle  \hat{x}_{k} \hat{x}_{m} +  \hat{x}_{m}\hat{x}_{k}\rangle}{2} \notag\\
&- \sum\limits_{j=1}^{2N}\sum\limits_{k=1}^{2N} \frac{\langle \hat{x}_{\ell}\hat{x}_{j}+ \hat{x}_{j}\hat{x}_{\ell}\rangle }{2} M_{jk} \Sigma_{km} \notag \\
 &+ \sum\limits_{j=1}^{2N}\sum\limits_{k=1}^{2N}\Sigma_{\ell j}\im\left(  C_{j}^{\ast} C_{k} \right) \frac{ \langle \hat{x}_{k} \hat{x}_{m}+\hat{x}_{m}\hat{x}_{k}\rangle}{2} \notag \\
 &+ \sum\limits_{j=1}^{2N}\sum\limits_{k=1}^{2N}\frac{ \langle \hat{x}_{\ell} \hat{x}_{j}+\hat{x}_{j}\hat{x}_{\ell} \rangle}{2} \im\left( C_{j}^{\ast}  C_{k} \right) \Sigma_{k m} \notag \\
& - \sum\limits_{j=1}^{2N}\sum\limits_{k=1}^{2N}  \Sigma_{\ell j} \re\left( C_{j}^{\ast}C_{k} \right) \Sigma_{k m} \notag \\
&-\sum\limits_{j=1}^{2N}\sum\limits_{k=1}^{2N}\Sigma_{\ell j} \bigg( M_{jk} + \im \left( C_{j}^{\ast}C_{k} \right) \bigg) \langle \hat{x}_{k} \rangle \langle\hat{x}_{m}\rangle  \notag \\
&- \sum\limits_{j=1}^{2N}\sum\limits_{k=1}^{2N}\langle\hat{x}_{\ell}\rangle \langle\hat{x}_{j}\rangle \bigg(  -M_{jk}   + \im \left( C_{j}^{\ast}C_{k} \right)\bigg) \Sigma_{km} \notag \\
=&\sum\limits_{j=1}^{2N}\sum\limits_{k=1}^{2N}\Sigma_{\ell j} \bigg( M_{jk} + \im \left( C_{j}^{\ast}C_{k} \right) \bigg) \bigg(\frac{  \langle  \hat{x}_{k} \hat{x}_{m} +  \hat{x}_{m}\hat{x}_{k}\rangle}{2} - \langle \hat{x}_{k} \rangle \langle\hat{x}_{m}\rangle \bigg) \notag  \\
& +\sum\limits_{j=1}^{2N}\sum\limits_{k=1}^{2N} \bigg( \frac{\langle \hat{x}_{\ell}\hat{x}_{j}+ \hat{x}_{j}\hat{x}_{\ell}\rangle }{2} - \langle\hat{x}_{\ell}\rangle \langle\hat{x}_{j}\rangle\bigg)\bigg(\im\left( C_{j}^{\ast}  C_{k} \right)-M_{jk}  \bigg) \Sigma_{km} \notag \\
& - \sum\limits_{j=1}^{2N}\sum\limits_{k=1}^{2N}  \Sigma_{\ell j} \re\left( C_{j}^{\ast}C_{k} \right) \Sigma_{k m} \notag \\
=&\sum\limits_{j=1}^{2N}\sum\limits_{k=1}^{2N}\Sigma_{\ell j} \bigg( M_{jk} + \im \left( C_{j}^{\ast}C_{k} \right) \bigg) V_{k m} \notag\\
&+\sum\limits_{j=1}^{2N}\sum\limits_{k=1}^{2N} V_{\ell j}\bigg(-M_{jk}+ \im\left( C_{j}^{\ast}  C_{k} \right) \bigg) \Sigma_{km} \notag \\
&- \sum\limits_{j=1}^{2N}\sum\limits_{k=1}^{2N}  \Sigma_{\ell j} \re\left( C_{j}^{\ast}C_{k} \right) \Sigma_{k m} \notag \\
=& \Sigma_{\ell :} \bigg( M + \im \left( C^{\dagger}C \right)\bigg) V_{: m} + V_{\ell :} \bigg(-M+ \im\left( C^{\dagger}  C \right) \bigg)\Sigma_{: m} \notag\\
&- \Sigma_{\ell :}\re\left( C^{\dagger}C \right) \Sigma_{: m},  \label{appendixA_proof_chapter1_covariance_evolution4}
\end{align}
where $V_{: m}$ denotes the $m$th column of $V$ and $V_{\ell :}$ denotes the $\ell$th row of $V$. It follows from \eqref{appendixA_proof_chapter1_covariance_evolution4} that 
\begin{align*}
\frac{dV}{dt} &= \Sigma\bigg( M + \im \left( C^{\dagger}C \right)\bigg) V + V \bigg(-M+ \im\left( C^{\dagger}  C \right) \bigg)\Sigma \\
&- \Sigma \re\left( C^{\dagger}C \right) \Sigma \\
&= \Sigma\bigg( M + \im \left( C^{\dagger}C \right)\bigg) V + V \bigg(M+ \im\left( C^{\dagger}  C \right) \bigg)^{\top}\Sigma ^{\top}\\
&+ \Sigma \re\left( C^{\dagger}C \right) \Sigma^{\top}. 
\end{align*}
That is, Equation~\eqref{chapter1_covfunction} holds. This completes the  derivation.\vspace{0.5cm}

\begin{rmk}\label{remark}
The above results can be easily extended to linear open quantum systems with multiple decoherence channels. 
By adding extra decoherence-induced terms (which are analogous to those obtained in~\eqref{appendixA_proof_chapter1_meanfunction_evolution2} and \eqref{appendixA_proof_chapter1_covariance_evolution2}), we can obtain the moment evolution equations for linear open quantum systems with multiple decoherence channels. The results are given by \eqref{chapter1_meanfunction} and \eqref{chapter1_covfunction}. 
\end{rmk}


\vspace{1cm}
\hspace{2.8cm}\textbf{Appendix: Notation}\hfill\\

 \begin{tabularx}{\columnwidth}{lX}
 \hline \\ [0.2ex]
$\mathbb{R}$, $\mathbb{R}^{m\times n}$          &Real numbers, real $m \times n$ matrices. \\ [1ex]
$\mathbb{C}$, $\mathbb{C}^{m \times n}$             &Complex numbers, complex $m\times n$ matrices. \\[1ex]
$I_{n}$  &The $n\times n$ identity matrix. \\[1ex]
$a^{\ast}$, $\hat{a}^{\ast}$ &The complex conjugate of a complex number $a$, the adjoint of an  operator $\hat{a}$.  \\[1ex]
$A^{\top}$ &Transpose of $A$, i.e.,  $\left(A^{\top}\right)_{jk}=A_{kj}$.  \\[1ex] 
$A^{\#}$ &Entrywise conjugate of $A$, i.e., $\left(A^{\#}\right)_{jk}=A_{jk}^{\ast}$.  \\[1ex] 
$A^{\dagger}$ &Conjugate transpose of $A$, i.e.,  $\left(A^{\dagger}\right)_{jk}=A_{kj}^{\ast}$.\\[1ex] 
$\re(A)$, $\im(A)$ &The real part, imaginary part of $A$, i.e., $\re(A)=\left(A+A^{\#}\right)/2$ and $\im(A)=\left(A-A^{\#}\right)/(2i)$.\\[1ex]
 $\tr(A)$  &Trace. \\[1ex]
\hline 
    \end{tabularx}
\end{document}